\documentclass[10pt]{article}
\date{}
\title{Qualms concerning Tsallis's condition of Pseudo-Additivity as a Definition of 
Non-Extensivity}
   \author{B. H. Lavenda$^1$ and J. Dunning-Davies$^2$\\
$^1$Universit\`a degli Studi  Camerino 62032 (MC) Italy;\\ email: bernard.lavenda@unicam.it\\
$^2$ Department of Physics, University of Hull, Hull HU6
7RX\\ England; email: j.dunning-davies@hull.ac.uk}
\usepackage{eufrak}

\newcommand{\summ}{\sum_{i=1}^m\,}

\begin{document}
\maketitle
\begin{abstract} 
The pseudo-additive relation that the Tsallis entropy satisfies has nothing whatsoever
to do with the super-and sub-additivity properties of the entropy. The latter properties,
like concavity and convexity, are couched in geometric inequalities and cannot be
reduced to equalities. Rather, the pseudo-additivity relation is a functional equation
that determines the functional forms of the random entropies. The Arimoto entropy 
satisfies a similar pseudo-additive relation and yet it is a first-order homogeneous
form. Hence, no conclusions can be drawn on the extensive nature of the system from either
the Tsallis or the Arimoto entropy based on the pseudo-additive functional equation.
\end{abstract}
\flushbottom
Tsallis \cite{Tsallis} has used the pseudo-additivity condition,
\begin{equation}
S_\alpha(A+B)=S_\alpha(A)+S_\alpha(B)+(1-\alpha)S_\alpha(A)S_\alpha(B) \label{eq:p-add}
\end{equation}
where we work in energy units in which $k=1$,
\begin{equation}
S_\alpha(P)=\frac{1-\summ\,p_i^\alpha}{\alpha-1} \label{eq:S-Tsallis}
\end{equation}
is the Tsallis entropy $\forall\;\alpha>0$, and the probability distribution 
$P=(p_1,\ldots,p_m)$ is complete. Supposedly \cite{Tsallis}, 
\begin{quote} $A$ and $B$ are two \emph{independent\/} systems in the sense that the 
probabilities of $A+B$ \emph{factorize\/} into those of $A$ and $B$ (i.e., $p_{ij}(A+B)
=p_i(A)p_j(B))$. We immediately see that, since in all cases $S_\alpha\ge0$ 
(\emph{nonnegativity\/} property), $\alpha<1$, $\alpha=1$ and $\alpha>1$ respectively
correspond to \emph{superadditivity\/} (\emph{superextensivity\/}), \emph{additivity\/}
(\emph{extensivity\/}) and \emph{subadditivity\/} (\emph{subextensivity\/}). 
\end{quote}
Yet, it appears odd that criteria  of super-and sub-additivity can be obtained through a
functional equation rather than as a geometric inequality as are the criteria for convexity
and concavity. The subadditive property is the \lq triangle inequality\rq\ 
\cite{BB}, whereas the geometrical interpretation of a concave function  is one that
never rises above its tangent plane at any
point. There are classes of functions which are defined by inequalities that are
weaker than convexity (concavity) and stronger than superadditivity (subadditivity)
\cite{Bruckner}.\par
To prove that (\ref{eq:S-Tsallis}) is always subadditive, consider the sum 
$\summ\,p_i^\alpha$. Companions to Minkowski's inequalities are \cite{Hardy} 
\begin{equation}
\summ\,(p_i+q_i)^\alpha>\summ\,p_i^\alpha+\summ\,q_i^\alpha \label{eq:super}
\end{equation}
for $\alpha>1$, and
\begin{equation}
\summ\,(p_i+q_i)^\alpha<\summ\,p_i^\alpha+\summ\,q_i^\alpha \label{eq:sub}
\end{equation}
for $0<\alpha<1$, where $Q=(q_1,\ldots,q_m)$ is another complete distribution. Inequality
(\ref{eq:super}) is the condition for superadditivity; the negative of which is subadditive.
Hence, for $\alpha>1$ the Tsallis entropy is subadditive. Inequality (\ref{eq:sub}) is the
criterion for subadditivity, and, hence, the Tsallis entropy is subadditive for $0<\alpha
<1$. This is in flagrant contradiction with the above conclusion that the entropy 
(\ref{eq:S-Tsallis}) is superadditive for $0<\alpha<1$ on the basis of (\ref{eq:p-add}).
\par
The pseudo-additivity relation (\ref{eq:p-add}) is actually expressed in terms of the
complete probability distributions
\begin{equation}
S_\alpha(PQ)=S_\alpha(P)+S_\alpha(Q)+(1-\alpha)S_\alpha(P)S_\alpha(Q).
\label{eq:p-add-T}
\end{equation}
The product of the two sums on the left-hand side does not imply 
additivity of two systems. Rather, (\ref{eq:p-add-T}) is a functional equation which
determines the form of the entropy.  Additive entropies
of degree-$1$ and degree-$\alpha$ can be derived from the first-
order homogeneous property of the weighted means \cite[p. 68]{Hardy}
\begin{equation}
\mathfrak{M}_{\varphi}(\lambda x)=\lambda\mathfrak{M}_{\varphi}(x),
\label{eq:homo}
\end{equation}
where $x$ stands for a set of discrete variables, $x_1,\ldots,x_m$. Since $\varphi$ is 
defined up to a constant, we may set
\begin{equation}
\varphi(1)=0. \label{eq:boundary}
\end{equation}
\par
Now (\ref{eq:homo}) stands for
\begin{equation}
\mathfrak{M}_{\varphi}(x)=\lambda^{-1}\mathfrak{M}_{\varphi}(
\lambda x)=\lambda^{-1}\varphi^{-1}\left(\summ\,
p_i\varphi(\lambda x)\right)=\mathfrak{M}_{\psi}(x). \label
{eq:mean-bis}
\end{equation}
On account of the translational invariancy of the weighted
mean
\begin{equation}
\psi(x)=a(\lambda)\varphi(x)+b(\lambda), \label{eq:linear}
\end{equation}
where $a\neq0$ and $b$ are functions of the parameter $\lambda$. 
Since it is apparent from (\ref{eq:mean-bis}) that $\psi(x)=
\varphi(\lambda x)$, we get
\begin{equation}
\varphi(\lambda x)=a(\lambda)\varphi(x)+b(\lambda) \label
{eq:1}
\end{equation}
However, according to (\ref{eq:boundary}) we must set $b(\lambda)
=\varphi(\lambda)$. Letting $\lambda$ become another positive 
variable, $y$,
(\ref{eq:linear}) becomes the functional equation
\begin{equation}
\varphi(xy)=a(y)\varphi(x)+\varphi(y). \label{eq:f1}
\end{equation}
By the law of associativity
\begin{equation}
\varphi(xy)=a(x)\varphi(y)+\varphi(x). \label{eq:f2}
\end{equation}

Setting the right hand sides of (\ref{eq:f1}) and (\ref{eq:f2}) equal to one another results in
\begin{equation}
\frac{a(x)-1}{\varphi(x)}=\frac{a(y)-1}
{\varphi(y)}=c \label{eq:c}
\end{equation}
where $c$ is a separation constant. Solving (\ref{eq:c})
 for $a(y)$ and substituting it into (\ref{eq:f1}) gives the
functional equation
\begin{equation}
\varphi(xy)=\varphi(x)+\varphi(y)+c\varphi(x)\varphi(y) 
\label{eq:add}
\end{equation}
which is known in information science as the additivity property 
of degree-$\alpha$.\par If $c=0$, it reduces to the classical equation
\[
\varphi(xy)=\varphi(x)+\varphi(y) \]
whose most general solution is $\varphi=A\log x$.\par If $x$ stands
for the probability $p$, then the best code length for an input 
symbol of probability $p$ is
\begin{equation} 
\varphi(p_i)=-\log p_i=n_i, \label{eq:cost}
\end{equation}
where $A=-1$ and $n_i$ is the length of sequence $i$. Then the average length,
$\summ\,p_in_i$ corresponds to the average entropy \cite{Campbell}
\begin{equation}
S=-\summ\,p_i\log p_i,
\label{eq:Shannon}
\end{equation}
which is the Shannon entropy, or additive entropy of degree-$1$.
\par
 With $c\neq0$, we insert $c\varphi(x)+1=g(x)$ into 
(\ref{eq:add}),
thereby reducing it to
\begin{equation}
g(xy)=g(x)g(y) \label{eq:classical}
\end{equation}
Since the general solution to the functional equation (\ref
{eq:classical}) is $g=x^r$, the random entropy function is
\begin{equation}
\varphi(p_i)=\frac{p_i^r-1}
{c} \label{eq:cost-bis}
\end{equation}
Since entropies must be positive, we must have $c<0$. Moreover,
 we want (\ref{eq:cost-bis}) to reduce to (\ref{eq:cost}) in 
the limit as $-c\rightarrow+0$ so that $r=-c$. Hence (\ref
{eq:cost-bis}) becomes
\begin{equation}
\varphi(p_i)=-\frac{p_i^{\alpha-1}-1
}{\alpha-1} \label{eq:cost-tris}
\end{equation}
where we have set $c=1-\alpha<0$. In the limit $\alpha\downarrow1$ it reduces to
(\ref{eq:cost}). The average of the entropy
(\ref{eq:cost-tris}) is the nonadditive entropy of degree-$\alpha$ \cite{Havrda,Daroczy,
Vajda,Aggarwal,Forte,Aczel,Mathai,Tsallis88} (\ref{eq:S-Tsallis})
which reduces to the additive entropy (\ref{eq:Shannon}) in the
limit as $\alpha\downarrow1$.\par
There is nothing unique about (\ref{eq:cost-tris}). If our only concern is that it
reduce to (\ref{eq:cost}) in the limit $\alpha\uparrow1$ then an equally likely candidate
is
\begin{equation}
\varphi(p_i)= \frac{1-p_i^{1-\alpha}}{1-\alpha}. \label{eq:cost-4}
\end{equation}
If we average (\ref{eq:cost-4}) with respect to $Q$, then
we must choose the $P$ in terms of $Q$ such that the new $P$ are normalized. In
terms of a variational problem we require
\[\summ q_i\frac{1-p_i^{1-\alpha}}{1-\alpha}+\mu\summ p_i=\mbox{min}.\]
where the constraint has been introduced using the Lagrange multiplier, $\mu$.
Performing the variation we get $p_i=(q_i/\mu)^{1/\alpha}$, and in order that  
$P$ be normalized
\[
p_i=\frac{q_i^{1/\alpha}}{\summ q_i^{1/\alpha}}. \]
Introducing these so-called \lq\lq escort\rq\rq\ probabilities \cite{Beck} back into 
(\ref{eq:cost-4}) and averaging give
\begin{equation}
S_\beta(Q)=\frac{\beta}{\beta-1}
\left\{1-\left(\summ q_i^{\beta}\right)^{1/\beta}\right\}, 
\label{eq:S-Arimoto}
\end{equation}
where we have set $\alpha=1/\beta$. The first-order homogeneous entropy
(\ref{eq:S-Arimoto}) appears to have been first introduced by Arimoto \cite{Arimoto}, 
and subsequently given a complete characterization by Boekee and van der Lubbe 
\cite{Lubbe}. \par
Just as the Tsallis entropy (\ref{eq:S-Tsallis}) obeys the pseudo-additivity relation
(\ref{eq:p-add-T}), the Arimoto entropy (\ref{eq:S-Arimoto})satisfies
\begin{equation}
S_\beta(PQ)=S_\beta(P)+S_\beta(Q)-\left(\frac{\beta-1}{\beta}\right)S_\beta(P)S_\beta
(Q). \label{eq:p-add-A}
\end{equation}
Following Tsallis' line of reasoning we would conclude from (\ref{eq:p-add-A}) that 
(\ref{eq:S-Arimoto}) is subadditive for $\beta>1$ and superadditive for $\beta<1$. 
However, according to Minkowski's inequalities \cite[p. 31]{Hardy}
\begin{equation}
\left(\summ\,(p_i+q_i)^\beta\right)^{1/\beta}<
\left(\summ\,p_i^\beta\right)^{1/\beta}+\left(\summ\,q_i^\beta\right)^{1/\beta}
\label{eq:sub-M}
\end{equation}
for $\beta>1$, and
\begin{equation}
\left(\summ\,(p_i+q_i)^\beta\right)^{1/\beta}>
\left(\summ\,p_i^\beta\right)^{1/\beta}+\left(\summ\,q_i^\beta\right)^{1/\beta}
\label{eq:super-M}
\end{equation}
for $\beta<1$. Thus, $\left(\summ\,p_i^\beta\right)^{1/\beta}$ is subadditive for $\beta
>1$ and superadditive for $\beta<1$. Since the negative of the former and 
positive of the latter is used to construct the Arimoto entropy, we conclude that it 
is always superadditive.\par
Moreover, concavity follows easily by setting $p_i=\lambda a_i$ and $q_i=(1-\lambda)
b_i$, where the sets of numbers $a_i$ and $b_i$ are nonnegative, and $0\le\lambda\le1$. Then inequality 
(\ref{eq:sub-M}) becomes the criterion for the convexity of 
$\left(\summ\,p_i^\beta\right)^{1/\beta}$,
\[\left(\summ\,(\lambda a_i+(1-\lambda)b_i)^\beta\right)^{1/\beta}<
\lambda\left(\summ\,a_i^\beta\right)^{1/\beta}+
(1-\lambda)\left(\summ\,b_i^\beta\right)^{1/\beta}\]
for $\beta>1$, and
\[\left(\summ\,(\lambda a_i+(1-\lambda)b_i)^\beta\right)^{1/\beta}>
\lambda\left(\summ\,a_i^\beta\right)^{1/\beta}+
(1-\lambda)\left(\summ\,b_i^\beta\right)^{1/\beta}\]
for $\beta<1$ is the condition
for its concavity. Since the Arimoto entropy takes the negative of the former  and  the
positive of the latter it is everywhere concave \cite{Lubbe}.\par
Likewise, the concavity of the Tsallis entropy (\ref{eq:S-Tsallis}) can be established 
from the observations that $\summ\,p_i^\alpha$ is convex for $\alpha>1$ and concave for
$0<\alpha<1$. Since the Tsallis entropy is formed from the negative of the former and the
positive of the latter, it is  concave for all $\alpha>0$.
\par
Furthermore, since the entropy is defined to within a constant, classically, the Arimoto
entropy (\ref{eq:S-Arimoto}) is a first-order homogeneous function, and, yet, it obeys the 
pseudo-additive relation (\ref{eq:p-add-A}), just like the Tsallis entropy 
(\ref{eq:S-Tsallis}) obeys the pseudo-additive relation (\ref{eq:p-add-T}). Hence, we can
conclude that this relation has nothing whatsoever to do with the extensivity properties,
or lack thereof, of the nonadditive entropies.\par

\end{document}